\documentclass[runningheads]{llncs}

\usepackage{listings}
\usepackage{xcolor}

\NewDocumentCommand{\codeword}{v}{%
\texttt{\textcolor{blue}{#1}}%
}

\definecolor{codegreen}{rgb}{0,0.6,0}
\definecolor{codegray}{rgb}{0.5,0.5,0.5}
\definecolor{codepurple}{rgb}{0.58,0,0.82}
\definecolor{backcolour}{rgb}{0.95,0.95,0.92}
\lstdefinestyle{mystyle}{
    backgroundcolor=\color{backcolour},
    commentstyle=\color{codegreen},
    keywordstyle=\color{magenta},
    numberstyle=\tiny\color{codegray},
    stringstyle=\color{codepurple},
    basicstyle=\ttfamily\footnotesize,
    breakatwhitespace=false,
    breaklines=true,
    captionpos=b,
    keepspaces=true,
    numbers=left,
    numbersep=5pt,
    showspaces=false,
    showstringspaces=false,
    showtabs=false,
    tabsize=2
}
\usepackage{hyperref}
\usepackage{graphicx}
\newcommand{\maps}{MaPS}

\begin{document}
\title{Predefined Software Environment Runtimes As A Measure For Reproducibility\thanks{supported by MaRDI, funded by the Deutsche Forschungsgemeinschaft (DFG), project number 460135501, NFDI 29/1 ``MaRDI - Mathematische Forschungsdateninitiative''.}}
\titlerunning{MaPS Runtimes As A Measure For Reproducibility}
\author{Aaruni Kaushik\orcidID{0009-0006-3061-6553}}
\authorrunning{A. Kaushik}
\institute{University of Kaieserslautern-Landau, DE \email{aaruni.kaushik@math.rptu.de}}

\maketitle              % typeset the header of the contribution
\begin{abstract}
Mathematical Research Data Initiative (MaRDI) is a consortium of the National Research Data Infrastructure (NFDI) aiming to bring
FAIR\footnote{FAIR Principles \url{https://www.go-fair.org/fair-principles/}} data practices to
mathematical research. In alignment with MaRDI, we have developed a way to preserve software 
packages into an easy to deploy and use sandbox environment we call a ``runtime'', via a
program we developed called \textbf{MaPS} : \textbf{Ma}RDI \textbf{P}ackaging \textbf{S}ystem
\cite{maps}. The program relies on Linux user namespaces to isolate a library environment from the
host system, making the sandboxed software reproducible on other systems, with minimal effort.
Moreover an overlay filesystem makes local edits persistent. This project will aid reproducibility
efforts of research papers: both mathematical and from other disciplines. As a proof of concept, we
provide runtimes for the OSCAR Computer Algebra System~\cite{oscar}, polymake software for research
in polyhedral geometry~\cite{polymake}, and VIBRANT Virus Identification By iteRative
ANnoTation~\cite{vibrant}. The software is in a prerelease state: the interface for creating,
deploying, and executing runtimes is final, and an interface for easily publishing runtimes is under
active development. We thus propose publishing predefined, distributable software environment
runtimes along with research papers in an effort to make research with software based results
reproducible.

\keywords{FAIR \and MaRDI \and Sandbox \and Runtime \and Reproducibility }
\end{abstract}

\section{Introduction}

As the capabilities of computer devices have grown, so have their use as tools in scientific
research. It is now possible to do wild and wonderful things by relying on computers to generate
useful results in the support of both research goals as well as arguments to reach them. However,
this also means that a computer program plays a more central role in the defence of a thesis. As
good science requires a peer review of the results, the computer programs so central to new
computational results must also be readily available to the reviewers. However, this is not
straightforward with software.

A naive approach to making computer software available is to simply make the source code available
to reviewers. But this poses problems with regards to ease of use, and also with regards to
reproducibility. It may not be the easiest thing to get running in the first place. This may just be
due to the convoluted way the prerequisites are meant to be installed. Or it may require obscure
versions of libraries which may conflict with existing libraries on a reviewer's host system. It
might just require a fixed version of a language which was the state of the art when the paper was
written, but has been deprecated and become unsupported on modern~computers. Thus, for a FAIR
approach to distributing research software, a system tailor-made for the job must be created. We
expect \maps~to be able to effectively preserve software in runtimes. Our work is heavily inspired
by the work done in this direction by Flatpak \cite{flatpak} for generic Linux software, and Valve
\cite{valve} for PC video games.

\subsection{Organization}

In Section 2, we describe a general overview of the system we have developed for packaging research
software in an easy to distribute and easy to use way. In Section 3, we describe in deep technical
detail how the system is implemented. Sections 4 and 5 provide some examples of using the system to
run packaged software.

A casual reader may skip Section 3, which is purely technical, and use sections 4 and 5 as a
rudimentary users manual. Section 3.4.4 may be of interest to users intending to create a package
runtime. For documentation of the project, the wiki\footnote{MaPS Wiki
\url{https://github.com/MaRDI4NFDI/maps/wiki}} should be consulted.

\section{MaRDI Packaging System: An Overview}

\maps~has only a very basic minimal set of requirements to work. It requires a reasonably recent
Linux kernel (5.11+ , released February 2021), and Python (3.8+, released October 2019)
pre-installed. Both of these are available in Ubuntu 22.04 LTS. The rest is grabbed automatically
during installation. Root access is only required at installation time for a system wide
installation, and resolving dependencies. \maps~can be installed with no root access when not
installing system wide. In this case, \maps~and any missing pre requisites must be installed
manually from source.

Instead of re-inventing the wheel, we make use of other free and open source\footnote{The
open source definition - Open Source Initiative \url{https://opensource.org/osd}} projects like
Bubblewrap~\cite{bwrap} and libostree~\cite{ostree} as mature, stable building blocks. This has a
two fold advantage: firstly, that we are saved from the workload of reimplementing parts of the
upstream work that we require, and secondly, that we inherit any improvements from these upstream
projects for free. While we trust that these projects will stick around, we have architectured the
solution in a way that in the unlikely scenario that these upstream projects are shut down, it is
possible to reimplement the functionality required from scratch. The building blocks are only a time
saving convenience.

In broad strokes, \maps~manages fetching and storing full runtimes from a network source onto the
local disk. It deploys these runtimes in a usable format from the storage provided by libostree, and
sets up the foundation of an overlay file system. When launching a runtime with MaPS, it dynamically
sets up an overlay layer on top of the published runtime, so that any local changes persist on disk,
without actually modifying the runtime. This allows end users to persist edits, while also leaving
open the option to easily reset back to the published state for perfect reproducibility.

\section{Technical Details}

\subsection{Bubblewrap}

Bubblewrap is a low level tool to create containers using user namespaces on Linux. It is also used
by other big projects like Flatpak and rpm-ostree. Using Bubblewrap, it is possible to create secure
isolated containers relatively easily, and with good granularity of the exact setup. Bubblewrap
prevents privilege escalation (see \cite[section System Security]{bwrap}),  and also adds a reaper
process into the container to avoid the docker PID 1 problem \cite{zombie}.

\subsection{libostree}

The libostree project (previously, just ostree) is a library for content addressed storage of
arbitrary filesystem trees. One can think of it as git, but for large directories instead of source
code. An ostree repository is initialized on the local disk, and a remote source is added, as a sort
of ``app store''. Data is then fetched from this remote using standard web protocols.

Along with managing the repository, and implementing the network activity, ostree also makes
efficient use of disk space. All files inside the repository are addressed by a hash of their
contents. Thus, files with identical data are stored only once on disk. Checking out trees from the
repository creates hardlinks on the local filesystem. Thus, file level deduplication is achieved.
However, note that any changes, even just a single bitflip, will defeat this deduplication.

\subsection{fuse-overlayfs}

The fuse-overlayfs project~\cite{fuse-overlayfs} is the userspace implementation of the
Overlayfs~\cite{overlayfs} filesystem. Overlayfs is a special Linux filesystem, which can present a
directory overlaid on top of another directory as a single merged filesystem, with any modifications
being written only to the ``upper'' layer. We need the userspace variant of this filesystem to
maintain the rootless nature of our application. It requires at minimum four directories: a ``lower
directory'', an ``upper direcotry'', a ``working directory'', and finally, a ``merged directory''.
With terminology borrowed from the physical analogy, an upper directory is overlaid on top of one or
more lower directories. The resulting tree is mounted onto the merged directory. A working directory
is required for temporary storage while the overlay is active.

Mounting a filesystem, even inside a container, presents a potential security risk, as the contents
of its superblock are executed in kernel mode. Thus, as of version 0.8.0, Bubblewrap does not
support mounting any filesystems in the container, not even Overlayfs. In principle, an overlay
operation is risk free as no superblocks are involved, and can safely be done as root inside a user
namespace in Linux 5.11+. However, this is not currently implemented in Bubblewrap. This point
serves as a first example of inheriting improvements for free: as native Overlayfs without root
access is a soon to be implemented feature on the roadmap for Bubblewrap, at which point, we can
stop depending on fuse-overlayfs, while still keeping all our functionality.

\subsection{MaPS}

At its core, \maps~is very simple and elegant. Its main task is to carefully orchestrate its
building blocks to acheive desired results. It tells ostree to set up a repository for all the data
in \texttt{\$XDG\_DATA\_HOME/org.mardi.maps/ostree/repo}, then checks out the files in
\texttt{\$HOME/.var/org.mardi.maps/<runtime\_name>}. Under this location, four subdirectories are
created: \texttt{rofs}, \texttt{rwfs}, \texttt{tmpfs}, and \texttt{live}. At the time of execution,
fuse-overlayfs is used to layer \texttt{rwfs} over \texttt{rofs}, and the result is exposed as a
merged writeable tree at \texttt{live}. Finally, Bubblewrap is used to create a user namespace using
\texttt{live} as the target filesystem tree. \texttt{\$HOME/Public} is passed through into the
namespace as \texttt{/home/runtime/Public}, and can be used to provide input, as well as extract
output from the runtime. When the runtime is exited, i.e., when the last process in the namespace
has exited, Bubblewrap will shut down the created namespace, and \maps~unmounts the filesytstem
created by fuse-overlayfs.

\subsubsection{Manifest File.}

Optionally, a runtime is allowed to contain a manifest file located at \texttt{/manifest.toml} in
the namespace. (Note that this is in runtime, which corresponds to
\texttt{<runtime>/live/manifest.toml} on the host.) This is a TOML\footnote{TOML: A config file
format for humans. \url{https://toml.io/en/}} file that can contain arbitrary metadata relating to
the runtime, and, more importantly, a custom command for \maps~invoke. This command is specified in
the \texttt{[Core]} section of the TOML file, as a string assigned to the variable \texttt{command}.
In the absence of this variable in the manifest file (or in the absence of a manifest file), a shell
is launched by default. This behaviour can be overridden by using the command line argument
\texttt{--command}. The Manifest file from the runtime
\texttt{org.oscar\_system.oscar/x86\_64/1.0.0} is included as an example:

\begin{lstlisting}
[Core]
command = "julia -J /tmp/jl_UuXQwY/Oscar.so --banner=no"

[Meta]
Project = "OSCAR -- Open Source Computer Algebra Research system, Version 1.0.0, The OSCAR Team, 2024. (https://www.oscar-system.org)"
URL = "https://www.oscar-system.org/"
\end{lstlisting}

\subsubsection{Extended Attributes.} libostree requires a file system with extended attributes
available and enabled to be able to manage a repository, and also to check out files from the
repository. This means that both required directories (\texttt{\$HOME} and
\texttt{\$XDG\_DATA\_HOME}) must be on a filesystem with extended attributes. In particular, this
means that NTFS disks or NFS shares will not work. If \texttt{\$HOME} is on such an unsupported
filesystem (as is common in large organizations), but access to another, supported, filesystem is
available, one may override these environment variables before invoking \maps~to point to locations
on the supported filesystem.

\subsubsection{Fakeroot.} Even though it appears that we have root access inside the user namespace
(the command \texttt{whoami} returns \texttt{root}), we don't really have access on the host.
This can lead to some conflict in operations like trying to \texttt{chown} a file. As root
in the namespace, the user expects to be able to do whatever they want in the namespace without an
error. But some commands like \texttt{chown} would need to actually change the owner of a file on
the host filesystem, which could be utilised to craft an attack to break out of the sandbox. While
\maps~does not try to be a security layer, such an operation is not allowed on the kernel level, and
results in an error. Fakeroot is a Linux utility which only simulates the result of such a command,
and returns success to the calling program. However, the kernel does not allow even this inside a
namespace, as a security measure. To get around this problem, fakeroot can be used with the
environment variable \texttt{FAKEROOTDONTTRYCHOWN} set to \texttt{1}. This makes fakeroot only
pretend to execute such commands, and return a success code to the calling program. This workaround
may be required in a \maps~runtime to successfully install new packages, or use \texttt{tar}.

\subsubsection{Creating Runtimes.} To create a runtime, start from a "minimal viable chroot"
environment. Debian rootfs is an excellent choice for reasons of compatibility and size. Initialise
the tree in a convenient location using \texttt{maps} \texttt{package} \texttt{--initialise}
\texttt{/path/to/tree}. Once initialised, open a live sandbox into the tree by running \texttt{maps}
\texttt{package} \texttt{--sandbox} \texttt{/path/to/repository}. Even though this is really a
sandbox inside the tree, it can be treated it as if one has SSH\footnote{Secure Shell
\url{https://en.wikipedia.org/wiki/Secure_Shell}}'d into a freshly installed debian box, and set up
programs as normal. When done, exit the sandbox, then package the runtime using \texttt{maps package
--commit name\_of\_repository/arch/version path/to/tree}. This will create a \maps~runtime out of
the tree.

\subsubsection{Publishing Runtimes.} A workflow for publishing runtimes created as in the previous
step is under active development. This will allow scientists to request their runtimes to be
published on the ``Official'' repository right from the \maps\ command line.

\subsection{Comparison with Competing Methods}

There are several methods for sharing a program for running on another machine ranging from sharing
just the source code, to docker containers (via a dockerfile), or a full fat Virtual Machine (VM)
disk image. We think \maps~is a superior option to these alternate methods. A \maps~runtime is more
complete than just sharing source code, more light weight than sharing a VM, and more streamlined
than running docker. A more in depth comparison is discussed in the full paper of this extended
abstract.

\subsection{Compatibility}

The technology enabling \maps~deeply depends on the features provided by the Linux kernel (isolation
via namespace). Technically, this means that the system and its benefits are limited to the Linux
kernel. As a result the program being packaged into a runtime \textbf{*MUST*} work on Linux!

Packaged runtimes provided by \maps~may still be used on systems powered by other kernels, via
virtual machines running Linux, or via other tightly integrated compatiblity layers. This is the
same strategy used by Docker on non Linux host OSs. Wrapping the compatibility layer inside the
\maps~program might be a consideration for a later date. For now, the recommended way of using
\maps~on Windows is via WSL\footnote{Windows Subsystem for Linux
\url{https://learn.microsoft.com/en-us/windows/wsl/}}, and on MacOS via lima\footnote{Lima: Linux
Machines \url{https://github.com/lima-vm/lima}}. More information is available on the
\maps~wiki\footnote{ MaPS Wiki \url{https://github.com/MaRDI4NFDI/maps/wiki/Non-Linux-OSs}}.

\section{Examples}

\subsection{AccurateArithmetic.jl}

The Julia package AccurateArithmetic.jl~\cite{papercorrectnessv3} implements the algorithms
described in \cite{accurate_arithmetic} in Julia. It also contains everything needed to reproduce
the results shown in the paper. Below is a full list of the actions to be taken by anyone wanting to
try and reproduce the results on their own Linux system (adapted for \maps~from
\cite[Appendix]{accurate_arithmetic})

Commands prefixed with a \codeword{sh>} prompt are to be entered in a shell; commands prefixed with
a \codeword{julia>} prompt are to be entered in a Julia interactive session. \maps~should have been
downloaded and installed beforehand by following the instructions at
\url{https://github.com/MaRDI4NFDI/maps/wiki/Installation}.

Be aware that step (3) in this procedure might take a few hours to complete. Afterwards, all
measurements should be available as JSON files in the \texttt{AccurateArithmetic.jl/test} directory.
After step (4), all figures showed in this paper should be available as PDF files in the same
directory.

\begin{enumerate}
    \item Get and launch the runtime:
    \begin{lstlisting}[language=Bash]
sh> maps -d org.juliamath.accuratearithmetic/x86_64/papercorrectnessv3
sh> maps -r org.juliamath.accuratearithmetic/x86_64/papercorrectnessv3\end{lstlisting}

    \item Start Julia and run the test suite:
    \begin{lstlisting}[language=Bash]
sh> cd /AccurateArithmetic.jl/test && julia --project
julia> using Pkg
julia> pkg"test"\end{lstlisting}

    \item Run the performance tests:
    \begin{lstlisting}[language=Bash]
# Additional dependencies for performance tests.
# Package testing fails if these are already included.
julia> pkg"add BenchmarkTools Plots Printf Statistics Test"
julia> exit()
sh> julia --project -O3 -L perftests.jl -e 'run_tests()'\end{lstlisting}

    \item Plot the graphs:
    \begin{lstlisting}[language=Bash]
sh> julia --project -L perfplorts.jl -e 'plot_results()'\end{lstlisting}
\end{enumerate}

\subsection{OSCAR}

OSCAR is a comprehensive open source computer algebra system for computations in algebra, geometry,
and number theory written in Julia\footnote{The Julia Programming Language
\url{https://julialang.org/}}. As a Julia program, it benefits from just in time compilation, with a
speedup step of precompilation. However, this precompilation step can be lengthy, and must be done
by each user \textbf{after} installation, upon first use. This can leave end users with the false
first impression of OSCAR being slow. As a way to mitigate this, OSCAR can be distributed as a
\maps~runtime, already pre-compiled, with a custom system image.

\begin{enumerate}
    \item Get the runtime:
    \begin{lstlisting}[language=Bash]
maps -d org.oscar_system.oscar/x86_64/latest\end{lstlisting}

    \item Start the runtime:
    \begin{lstlisting}[language=Bash]
maps -r org.oscar_system.oscar/x86_64/latest\end{lstlisting}
\end{enumerate}

Starting the runtime will automatically also start a Julia REPL\footnote{Read-evaluate-print loop}
environment and load a precompiled copy of OSCAR. To update the runtime, for the latest version of
OSCAR, execute

\begin{lstlisting}
maps --update org.oscar_system.oscar/x86_64/latest\end{lstlisting}

\section{Beyond Mathematical Software}

We present an example which showcases non mathematical software which can benefit from MaPS.

\subsection{VIBRANT}

VIBRANT is a tool for automated recovery and annotation of bacterial and archaeal viruses,
determination of genome completeness, and characterization of viral community function from
metagenomic assemblies. VIBRANT uses neural networks of protein annotation signatures and genomic
features to maximize identification of highly diverse partial or complete viral genomes as well as
excise integrated proviruses. VIBRANT is a novel and useful tool which, unfortunately, requires
versions of software which are not easy to install in modern operating systems. This presents a
perfect opportunity for research software to be ported into a \maps~runtime, such that it can be run
on current systems without having to install outdated versions of software, which might cause
conflicts with, and threaten the stability of an up to date system.

As before, installing it is as simple as calling the \maps~deploy function with the runtime name :

\begin{lstlisting}
maps --deploy github.anantharaman.vibrant/x86_64/1.2.1\end{lstlisting}

\section{Concluding Remarks}

Packaging software for distribution always means an overhead in terms of work for a researcher.
However, correctly doing so leads to great progress in the reproducibility and thus the credibility
of the scientific work at hand. It also makes the research more FAIR. We have developed a software
tool to simplify this packaging process, keeping in mind the long term reproducibility of the work.
The runtimes we create are lighter than a VM, more straighforward than docker, and more complete
than just a source distribution.

\bibliographystyle{splncs04}
\bibliography{mybibliography}

\begin{thebibliography}{10}
\providecommand{\url}[1]{\texttt{#1}}
\providecommand{\urlprefix}{URL }
\providecommand{\doi}[1]{https://doi.org/#1}

\bibitem{overlayfs}
Overlay filesystem - the linux kernel documentation (2014), \url{https://www.kernel.org/doc/html/latest/filesystems/overlayfs.html?highlight=overlayfs}

\bibitem{flatpak}
Flatpak: The future of apps on linux (2015), \url{https://flatpak.org/}

\bibitem{bwrap}
Bubblewrap - low-level unprivileged sandboxing tool used by flatpak and similar projects (2016), \url{https://github.com/containers/bubblewrap}

\bibitem{ostree}
libostree - operating system and container binary deployment and upgrades (2016), \url{https://github.com/ostreedev/ostree}

\bibitem{papercorrectnessv3}
Accurate{A}rithmetic.jl - calculate with error-free, faithful, and compensated transforms and extended significands (2019), \url{https://github.com/JuliaMath/AccurateArithmetic.jl/tree/paper-correctness-2019}

\bibitem{fuse-overlayfs}
fuse-overlayfs - fuse implementation for overlayfs (2020), \url{https://github.com/containers/fuse-overlayfs}

\bibitem{valve}
Valvesoftware/steam-runtime - a runtime environment for steam applications (2020), \url{https://github.com/ValveSoftware/steam-runtime}

\bibitem{accurate_arithmetic}
Elrod, C., F{\'e}votte, F.: {Accurate and Efficiently Vectorized Sums and Dot Products in Julia} (Aug 2019), \url{https://hal.science/hal-02265534}, version submitted to the Correctness2019 workshop

\bibitem{polymake}
Gawrilow, E., Joswig, M.: {\tt polymake}: a framework for analyzing convex polytopes. In: Polytopes---combinatorics and computation ({O}berwolfach, 1997), DMV Sem., vol.~29, pp. 43--73. Birkh\"auser, Basel (2000)

\bibitem{maps}
Kaushik, A.: Mardi packaging system (2023), \url{https://github.com/MaRDI4NFDI/maps}

\bibitem{vibrant}
Kieft, K., Zhou, Z., Anantharaman, K.: Vibrant: Virus identification by iterative annotation copyright (2020), \url{https://github.com/AnantharamanLab/VIBRANT}

\bibitem{zombie}
Lai, H.: Docker and the pid 1 zombie reaping problem (2015), \url{https://web.archive.org/web/20240228145942/https://blog.phusion.nl/2015/01/20/docker-and-the-pid-1-zombie-reaping-problem}

\bibitem{oscar}
Oscar -- open source computer algebra research system (2024), \url{https://www.oscar-system.org}

\end{thebibliography}
\end{document}